\documentclass{article}
\usepackage[utf8]{inputenc}
\usepackage{graphicx}
\usepackage[portrait, margin=0.8in]{geometry}
\usepackage{hyperref}
\usepackage{authblk}
\usepackage[braket, qm]{qcircuit}
\usepackage{listings}
\usepackage{todonotes}
\usepackage{enumitem}
\usepackage{float}
\usepackage{siunitx}
\usepackage{adjustbox}
\usepackage{amsmath,amsthm,amssymb}
\usepackage{comment}
\usepackage{enumitem} 
\usepackage[numbers,sort&compress]{natbib}
\title{Digital Zero-Noise Extrapolation with Quantum Circuit Unoptimization}

\author[1]{Elijah Pelofske\thanks{Email: epelofske@lanl.gov}}
\author[2]{Vincent Russo\thanks{Email: vincent@unitary.foundation}}
\affil[1]{Los Alamos National Laboratory, Information Systems \& Modeling}
\affil[2]{Unitary Foundation}

\date{\vspace{-6ex}}

\begin{document}
\maketitle

\begin{abstract}

Quantum circuit unoptimization is an algorithm that transforms a quantum circuit into a different circuit that uses more gate operations while maintaining the same unitary transformation. We demonstrate that this method can implement digital zero-noise extrapolation (ZNE), a quantum error mitigation technique. By employing quantum circuit unoptimization as a form of circuit folding, noise can be systematically amplified. The key advantages of this approach are twofold. First, its ability to generate an exponentially increasing number of distinct circuit variants as the noise level is amplified, which allows noise averaging over many circuit variants with slightly different circuit structure. Averaging over these variants can mitigate the effect of biased error propagation due to the significantly altered circuit structure from quantum circuit unoptimization, or biased noise sources on a quantum processor. Second, quantum circuit unoptimization by design resists circuit simplification back to the original unmodified circuit, making it plausible to use ZNE in contexts where circuit compiler optimization is applied server-side. We evaluate the effectiveness of quantum circuit unoptimization as a noise-scaling method for ZNE in two test cases using depolarizing noise numerical simulations: random quantum volume circuits, where the observable is the heavy output probability, and QAOA circuits for the (unweighted) maximum cut problem on random 3-regular graphs, where the observable is the cut value. We show that using quantum circuit unoptimization to perform ZNE can approximately recover signal from noisy quantum simulations. 

\end{abstract}

\section{Introduction}
\label{section:introduction}

Quantum error mitigation is a class of algorithms and techniques whose goal is
to suppress errors in quantum computations on near-term noisy quantum
computers~\cite{cai2023quantum, li2017efficient, temme2017error,
endo2018practical, kandala2019error, strikis2021learning, giurgica2020digital,
larose2022mitiq, endo2021hybrid, kim2023scalable, koczor2021exponential,
he2020zero, huggins2021virtual, pascuzzi2022computationally, song2019quantum}.
Generally, error suppression can refer to techniques that do not incur any
computational overhead, such as dynamical decoupling~\cite{viola1999dynamical,
souza2012robust, facchi2004unification, santos2005dynamical, viola2005random,
pokharel2018demonstration, sekatski2016dynamical}. Of the techniques that
require classical post-processing, there are several families of techniques that
have been proposed, all of which typically refer to quantum error mitigation.
Almost all post-processing-based quantum error mitigation strategies require a
real number observable quantity to be measured from the samples obtained by the
quantum computer---with some exceptions that use sampling-based methods and
knowledge of the error model of the quantum hardware to also extract
error-mitigated samples~\cite{barron2024provable}. In general, however, the
resources required to implement quantum error mitigation have an exponential
overhead~\cite{takagi2022fundamental, quek2024exponentially}---and therefore
these techniques are always a type of heuristic approximation and will likely
not scale to large quantum circuit sizes due to their resource requirement
scaling, but can improve the signal measured from current noisy quantum
computers. 

Numerous quantum error mitigation algorithms have been developed, including
probabilistic error cancellation (PEC)~\cite{mari2021extending,
van2023probabilistic, ma2024limitations}, Clifford data regression
(CDR)~\cite{czarnik2021error}, and machine learning-based
approaches~\cite{strikis2021learning, sack2024large, liao2024machine,
kim2020quantum, bennewitz2022neural}. This study focuses on zero-noise
extrapolation (ZNE)~\cite{giurgica2020digital, he2020zero,
pascuzzi2022computationally, kandala2019error, mari2021extending,
temme2017error}, a quantum error mitigation technique that amplifies the noise
in a quantum computation by a specified factor. The resulting measurements
contain a controlled amount of noise or error. An observable quantity is
extracted from these measurements, and an extrapolation is performed to estimate
the zero-noise limit. ZNE, in comparison to many other quantum error mitigation
algorithms, has a lower computational and sampling overhead, making it more
accessible to use on current quantum computers. To summarize, ZNE is comprised
of the following components:

\begin{enumerate}[noitemsep]
    \item A noise amplification method. This can be done in various ways, but
    the required property is that the quantum circuit is executed with a
    quantifiable amount of noise introduced into the computation. 
    \item The computation of interest must have a numerical quantity, an
    observable, that can be extracted from the measured qubit states (i.e.,
    samples). 
    \item A regression curve fits the noise-amplified observable data and then
    can be used to extrapolate that noise observable fit to the zero-noise
    point. 
\end{enumerate}

Generally, there are two methods for scaling the noise. The first is digital,
meaning in the circuit representation of the unitary that is being
simulated---typically, this comes in the form of an increased number of circuit
instructions that are still implementing an identical unitary (under noiseless
conditions). The second is by stretching the pulse-level waveforms on the
hardware~\cite{kandala2019error, kim2023evidence} that are implementing the
circuit-level instructions---in particular, by still implementing the same
operations, but over longer timescales, which thus creates more noise due to the
limited qubit coherence times and an accumulation of errors over time. 

In this study, we demonstrate that digital zero-noise extrapolation,
specifically, the noise amplification portion of ZNE, can be implemented using
quantum circuit unoptimization, which was introduced in~\cite{mori2024quantum}
to test the effectiveness of quantum circuit compiler optimization. Quantum
circuit unoptimization has several valuable properties compared to existing
digital noise amplification methods. 

First, quantum circuit unoptimization enables the generation of an exponentially
growing number of potential circuit variants as the circuit noise is amplified.
Circuit variant here means a circuit with a different set of elementary quantum
gates, but still implements exactly the same unitary operation. In contrast,
other methods, such as global unitary circuit folding, generate exactly one
circuit per noise scale factor. This exponential growth in possible circuit
variants can be particularly advantageous when the noise profile of the quantum
hardware is highly variable or non-uniform. By averaging over multiple circuit
variants---each representing a slightly different implementation of the same
logical unitary operation---one can mitigate the impact of localized noise
effects. This circuit variant averaging also mitigates noise biasing that can
occur from the inherent differences in quantum circuit structure from the
unoptimization routine. This approach reduces bias in the measured observables,
as the diverse noise profiles, or specifically physical implementations, of the
circuit effectively ``sample'' different noise environments, leading to more
robust statistics.

Second, the unoptimization routines of~\cite{mori2024quantum} are designed to
resist removal by compiler optimization passes in standard quantum compilers,
such as Qiskit, at least with current versions. Compiler optimization often aims
to simplify quantum circuits by eliminating redundant operations or reducing
gate counts, which can inadvertently reverse noise amplification steps in
techniques like unitary folding. Quantum circuit unoptimization circumvents this
issue by introducing structural complexity that compilers do not recognize or
simplify. As a result, the noise amplification achieved through quantum circuit
unoptimization is more resilient to server-side circuit optimization, ensuring
that the intended scaling remains intact during execution.

Third, quantum circuit unoptimization allows for fractional noise scale factors.
While this is not a property unique to quantum circuit unoptimization, it is a
significant advantage for ZNE. Fractional noise scaling enables finer control
over the noise amplification process, particularly for scale factors close to
$1.0$. This capability is especially valuable in highly noisy computation
scenarios where slight deviations from the original circuit may already
introduce significant noise. Quantum circuit unoptimization facilitates accurate
extrapolation to the zero-noise limit by enabling precise control over noise
scaling.

\section{Methods}
\label{section:methods}

\subsection{Quantum Circuit Unoptimization}
\label{section:methods_quantum_circuit_unoptimization}

The primary algorithm employed in this study is \emph{quantum circuit
unoptimization}, a novel quantum circuit compiler algorithm introduced
in~\cite{mori2024quantum}. This algorithm systematically transforms a given
quantum circuit into a more complex version while preserving its unitary
equivalence under ideal, noiseless conditions. By doing so, quantum circuit
unoptimization provides a mechanism for controlled noise amplification. The
unoptimization process follows a four-step iterative ``recipe'' that is applied
repeatedly to the input quantum circuit. Each iteration increases the circuit's
depth and gate count, introducing additional opportunities for noise while
preserving the original computation.

At a high level, for a given quantum circuit, the recipe steps are:
\begin{enumerate}[noitemsep]
    \item Insert: Insert a two-qubit gate $A$ and its Hermitian conjugate
    $A^{\dagger}$ between two 2-qubit gates in the circuit denoted as $B_1$ and
    $B_2$. We choose $A$ such that it is a random two-qubit unitary. 
    \item Swap: Swap the $B_1$ gate with the $A^{\dagger}$ gate in the circuit,
    replacing $A^{\dagger}$ with $\widetilde{A^\dagger}$.
    \item Decompose: Decompose multi-qubit unitary gates into elementary gates.
    \item Synthesize: Apply quantum circuit synthesis.
\end{enumerate}
For the insert step, a random two-qubit unitary $A$ and its Hermitian conjugate
$A^{\dag}$ are introduced into the circuit. These gates are inserted between two
existing two-qubit gates in the circuit, denoted as $B_1$ and $B_2$. The
insertion of $A$ and $A^{\dag}$ does not alter the logical functionality of the
circuit, as the combination $A^{\dag}A = I$. The selection of $B_1$ and $B_2$
can follow different strategies, such as choosing gates that share a common
qubit (the \emph{concatenated strategy}) or selecting them randomly (the
\emph{random strategy}). The choice of strategy affects the unoptimization
procedure's characteristics and its impact on circuit structure. The
concatenated strategy, as we implement it, is specifically a deterministic two
2-qubit gate selection procedure, where we choose the first pair of two-qubit
gates that share a common qubit. This means specifically that for the
concatenated strategy, there is no randomness in the selection of this pair of
$2$-qubit gates. However, one could certainly utilize random selection of pairs
of two-qubit gates that share a common qubit. 

After insertion, the $B_1$ gate is swapped with the $A^{\dag}$ gate. This
swapping operation introduces a new gate, denoted as $\widetilde{A^{\dag}}$,
which encapsulates the interaction between $B_1$ and $A^{\dag}$. Specifically,
\begin{equation}
    \widetilde{A^{\dag}} = 
    \left(I \otimes B_1\right)^{\dag} 
    \left(I \otimes A^{\dag}\right) 
    \left(B_1 \otimes I\right),
\end{equation}
representing the transformation of $A^{\dag}$ under the influence of $B_1$. This
step introduces additional complexity in the circuit by altering the placement
and the nature of the gates while maintaining unitary equivalence.

\begin{figure*}[ht!]
    \centering
    \includegraphics[width=\linewidth]{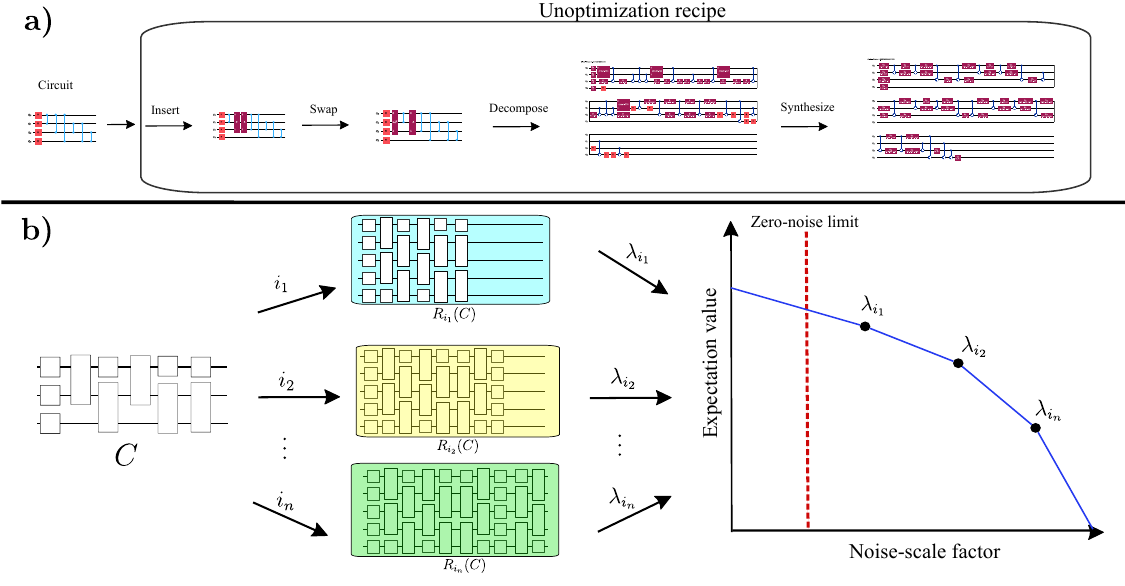}
    \caption{Depiction of quantum circuit unoptimization applied to ZNE. Panel
    \textbf{a)} shows a small, but representative, example of quantum circuit
    unoptimization being used. The example is a 4-qubit fully connected graph
    state circuit which is the input for a single round of the circuit
    unoptimization algorithm, using the concatenated pair of two-qubit gates
    selection strategy (the common qubit is $q_0$). Unoptimization consists of
    gate insert, swap, decompose, and synthesis operations. The right-most
    circuit shows the final state of the original circuit after a single round
    of unoptimization has been applied to it. The input and resulting circuits
    are unitarily equivalent to one another, which remains true even after
    repeated rounds of unoptimization. Recursive application of quantum circuit
    unoptimization results in a growing combinatorial explosion of possible
    circuit (equivalent) variants with different circuit structure as more
    iterations are applied. This large number of potential circuit variants is
    due to random choice of two 2-qubit gates during the Insert step (this is
    only true for the random strategy), and because the two-qubit gate $A$ in
    the Insert step is chosen as a random unitary. Measurement gates are not
    shown for brevity in the circuit renderings, but in this case the state of
    all 4 qubits at the end of the circuit are measured. Additional information
    about the gates used in the Qiskit-generated circuit diagrams is provided
    in the Appendix~\ref{section:appendix}. Panel \textbf{b)} shows a conceptual
    diagram of using quantum circuit unoptimization to amplify noise and then
    carry out zero noise extrapolation, given an arbitrary input circuit $C$
    consisting of single and two-qubit gates. For some choice of integers $i_1 <
    i_2 < \cdots < i_n$, we apply $i_k$ iterations of the recipe to the circuit
    $C$, denoted as $R_{i_k}(C)$ for some $k \in \{1, \ldots, n\}$. For each
    resulting circuit from $R_{i_k}(C)$, we obtain a corresponding noise-scale
    factor as a function of the resulting circuit's depth and the original
    circuit's depth. The greater the number of recipe iterations corresponds to
    a circuit with increasing depth (and hence increased noise). We compute the
    expectation value of each noise-scale factor and extrapolate to the
    zero-noise limit via some extrapolation method (e.g., linear, Richardson,
    exponential, etc.).}
    \label{fig:unopt}
\end{figure*}

Now containing additional gates from the first two steps, the circuit undergoes
a decomposition and gate synthesis process. These last two steps can be thought
of as effectively grouped together because this is just a circuit compilation
routine to compile down to some quantum instruction gateset. Critically, these
steps are expressing the unoptimized circuit, which is now comprised of
Multi-qubit unitary gates, including $\widetilde{A^{\dag}}$, and are broken down
into elementary gates, such as single-qubit rotations and CNOT gates. This
decomposition is performed using Qiskit~\cite{javadiabhari2024quantum},
specifically a single decomposition pass. The final step optimizes the
decomposed circuit using a synthesis procedure, in this case, we implement this
step in the form of the Qiskit transpiler~\cite{javadiabhari2024quantum}, with
optimization level $3$ (which is the maximum), targeting a particular gateset.
If this circuit is being implemented on a specific quantum computing
architecture, then the native gateset of that quantum computer would be used as
the target gateset when performing these decomposition and synthesis steps.

The unoptimization recipe can be applied iteratively to a quantum circuit with
each application subsequently increasing the circuit's depth, increasing the
total number of instructions and altering the overall circuit structure. Let
$R(C)$ represents the application of the unoptimization recipe to a circuit $C$.
After $i$ iterations, the resulting circuit can be expressed recursively as
\begin{equation} 
    R_i(C) := R(R(\dots R(C) \dots)),
\end{equation}
where the recipe is applied $i$ times. Each application of the recipe
corresponds to a noise scale factor, $\lambda_i$, which quantifies the ratio of
the depth of the original circuit to the depth of the circuit after $i$
iterations. Formally, the noise scale factor is defined as
\begin{equation} 
    \lambda_i = \frac{n(C)}{n(R_i(C))},
    \label{equation:noise_scale_factor}
\end{equation}
where $n(C)$ and $n(R_i(C))$ denote the total number of single- and two-qubit
gates in the original and the $i$-th unoptimized circuits, respectively. This
metric, rather than circuit depth alone, provides a more comprehensive measure
of circuit complexity and the opportunities for noise to accumulate.
Importantly, eq.~\eqref{equation:noise_scale_factor} is only an approximation of
the true noise amplification because it does not capture the differences in the
circuit structure and error propagation. We use this metric because the circuit
depth ratio corresponds to noise amplification and is easy to compute, whereas
comprehensive error propagation measurement can be computationally intensive. 

Quantum circuit unoptimization is used in this study as the noise-scaling
mechanism within the zero-noise extrapolation (ZNE) error mitigation framework.
The process is depicted in Figure~\ref{fig:unopt}, using an example input
circuit of a 4-qubit graph state. After applying the unoptimization recipe
iteratively to generate circuits with increasing noise levels, the resulting
circuits are executed on a quantum device or simulator. For each circuit
$R_i(C)$, the expectation value of a target observable is measured. The
corresponding noise scale factor $\lambda_i$ is associated with each
measurement. Using these data points, an extrapolation method (e.g., linear,
Richardson, or exponential extrapolation) is applied to estimate the zero-noise
limit of the observable. 

\subsection{Test Cases}
\label{section:methods_test_cases}

We evaluate quantum circuit unoptimization for ZNE using two different classes
of quantum circuits. The first class consists of quantum volume (QV) circuits,
where the observable is the heavy output probability (HOP). Quantum volume is a
benchmark for noisy quantum computers that requires coherent sampling of dense,
random quantum circuits of a specific form~\cite{cross2019validating,
baldwin2022reexamining, pelofske2022quantum, jurcevic2021demonstration}, where
the goal of the benchmark is that the noisy quantum computer must reliably
sample a quantum circuit that has approximately the same depth (circuit depth)
as its width (number of qubits being used in the circuit). ZNE was previously
applied to quantum volume in a technique known as \emph{effective quantum
volume}~\cite{pelofske2024increasing, larose2022error}, which extends the
quantum volume benchmark to include reliable estimation of noise mitigated
quantities when the circuit is executed on a noisy quantum computer. 

QV circuits are constructed such that the number of layers is equal to the
number of qubits, creating random dense circuits with an inherent symmetry in
depth and width. These circuits exhibit a key property: under ideal conditions,
they concentrate amplitude into a subset of computational basis states, leading
to a heavy output probability of approximately $\frac{1 + \ln(2)}{2} \approx
0.85$ in the asymptotic limit~\cite{aaronson2017complexity}. However, as
decoherence and noise accumulate, this amplitude concentration diminishes, and
the heavy output probability approaches $0.5$, corresponding to uniform random 
sampling. Therefore, the heavy output probability measure is defined over the range
$[0, 1]$.

The intuition behind the QV benchmark is that it requires the noisy quantum
hardware to be able to sample dense circuits that have the same number of qubits
as the gate depth is---meaning that the benchmark does not disproportionately
favor very shallow-depth circuits or very high-depth circuits acting on a few
qubits. Quantum volume, however, requires full classical state-vector simulation
of each circuit, meaning that this is not scalable into a regime where full
classical simulation of the quantum system can not be performed. 

The second class of quantum circuits we evaluate consists of those used in the
Quantum Approximate Optimization Algorithm (QAOA)~\cite{farhi2014quantum,
farhi2015quantum}, specifically applied to the Maximum Cut (Max-Cut) problem on
random 3-regular graphs~\cite{harrigan2021quantum, zhou2020quantum,
basso2021quantum, wang2018quantum, crooks2018performance, guerreschi2019qaoa,
marwaha2021local, hastings2019classical}. Max-Cut is a well-known NP-hard
combinatorial optimization problem~\cite{goemans1995improved, trevisan2009max,
festa2002randomized}. 

QAOA circuits consist of alternating simulations of layers of a cost Hamiltonian $H_C$, called the phase separator, (which is derived from the problem's objective
function) and a mixer Hamiltonian $H_M$. The mixer gives parameterized interference, or state transitions, between solutions of different cost value. The number of layers, denoted
as $p$, determines the depth of the circuit. The two Hamiltonians $H_C$ and $H_M$ do
not commute, and each is parameterized by a set of real numbers $\vec{\gamma}$
and $\vec{\beta}$, respectively, referred to as \emph{QAOA angles}.

QAOA aims to approximate ground state(s) of diagonal (classical) cost Hamiltonians
corresponding to a combinatorial optimization problem. However, for QAOA to
perform well, a sufficiently large depth $p$ is required, and the QAOA
angles must be optimized. These parameters are typically optimized
variationally, or designed to approximate adiabatic quantum
evolution~\cite{sack2021quantum}. In this study, we use the fixed-angle QAOA
parameters proposed in~\cite{wurtz2021fixed}, which are known to perform well on
random 3-regular Max-Cut instances due to the parameter concentration property
of the QAOA control parameters, which have been seen in numerous problem
classes~\cite{akshay2021parameter, brandao2018for, basso2022quantum}. The
observable of interest in our ZNE simulations is the expected graph cut value
given by the Hamiltonian,
\begin{equation}
    H(z) = \sum_{(u, v) \in E} z_u z_v
    \label{equation:maxcut_ising}
\end{equation}
where $E$ is the edge set defining the interactions in the graph $G = (V, E)$,
and $z$ is a spin configuration represented as a vector of values $z \in \{-1,
+1\}^n$. The minimum energy of this Hamiltonian corresponds to the optimal Max-Cut
solution.

Applying quantum circuit unoptimization to QAOA circuits is a compelling test
case for several reasons. First, the noise characteristics of QAOA circuits are
well-studied, and the cost Hamiltonian's expectation value is sensitive to
noise, making it a suitable benchmark for evaluating ZNE. Second, the structure
of QAOA circuits, which involves specific patterns of entangling gates and
parameterized rotations, allows us to investigate how unoptimization interacts
with structured quantum algorithms. Third, as QAOA circuits are actively
explored for practical quantum applications, demonstrating the effectiveness of
unoptimization as a noise-scaling technique in this context could be relevant to
broader algorithmic developments.

\begin{figure*}[ht!]
    \centering
    \includegraphics[width=0.49\linewidth]{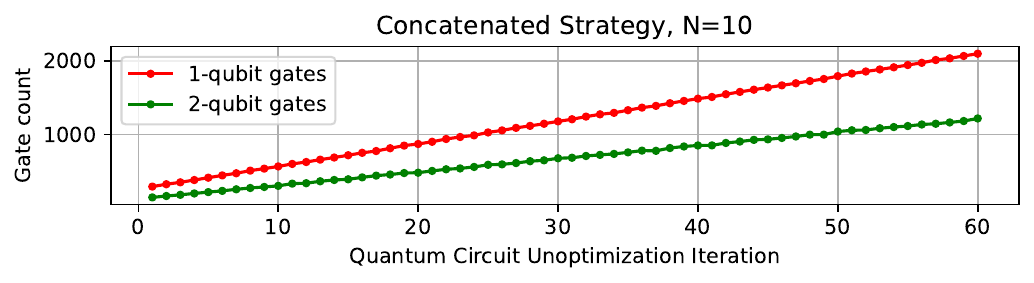}
    \includegraphics[width=0.49\linewidth]{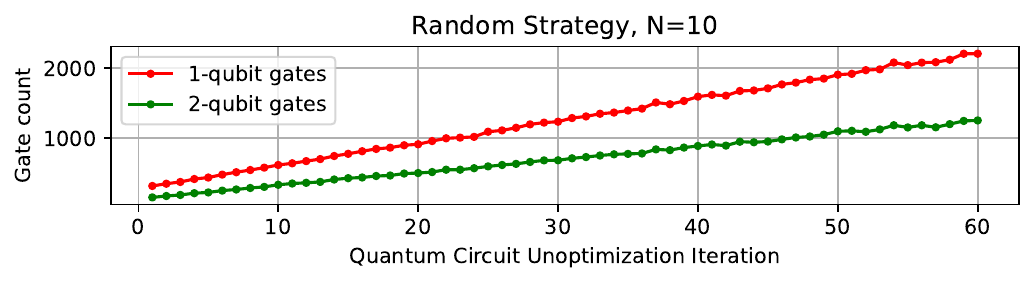}
    \includegraphics[width=0.49\linewidth]{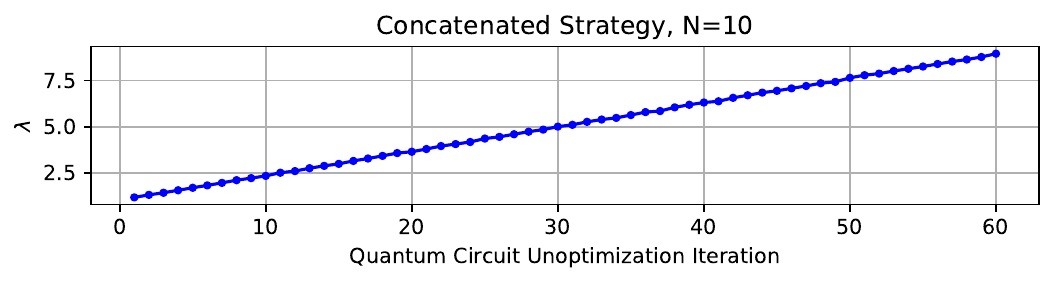}
    \includegraphics[width=0.49\linewidth]{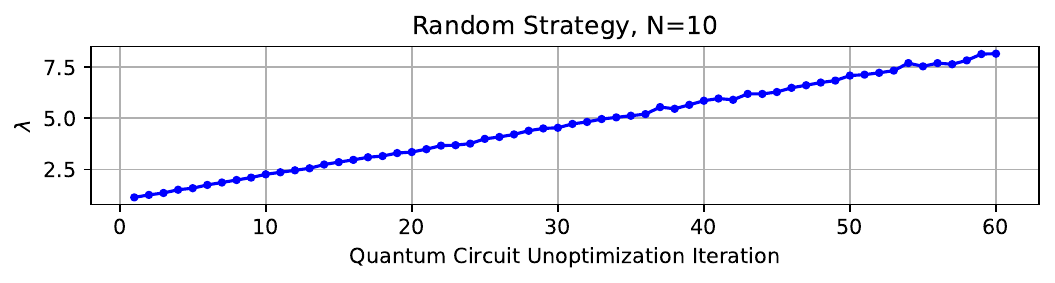}
    \caption{Quantum circuit unoptimization scaling as a function of recursive
    iteration number (x-axis) applied to a $10$ qubit quantum volume circuit.
    The top row shows single and two-qubit gate counts. The bottom row shows the
    overall circuit noise scale factor $\lambda$. The left-hand column uses a
    two-qubit gate selection based on a shared qubit index (concatenated
    strategy), and the right-hand column selects pairs of two-qubit gates at
    random (random strategy). The two-qubit gate selection is based on a
    shared qubit shows a more stable noise scale factor scaling.}
    \label{fig:circuit_size_scaling}
\end{figure*}

\subsection{Numerical Simulations}
\label{section:methods_numerical_simulations}

We present several numerical simulations to analyze the noise characteristics of
using quantum circuit unoptimization for ZNE. All numerical simulations are
performed using depolarizing noise (error) models using the Qiskit quantum
programming library~\cite{javadiabhari2024quantum} (version number
\texttt{1.3.2}). Specifically, a depolarizing error rate of $0.001$ is applied
to all single and two-qubit gates, and a finite shot count of $1\mathrm{e}{-6}$
measurements are used for each circuit and noise scale factor execution. The
quantum circuit unoptimization noise scale factor
(eq.~\eqref{equation:noise_scale_factor}) is computed after applying the Qiskit
transpiler optimization pass to each unoptimization iteration. The
unoptimization iteration is applied to the original circuit after applying the
Qiskit transpiler optimization pass. \texttt{optimization\_level} $3$ was used
for all Qiskit optimization passes. All circuits are transpiled to the basis
gates of \texttt{cx} (CNOT) and \texttt{U3} -- reported single and two-qubit
gate counts are based on only these two gates. 

The extrapolation functions we test are linear and quadratic functions. The
coefficients for these extrapolations are fit using least squares curve fitting.
The quadratic function is $ax^2 +bx +c$ where $a, b$, and $c$ are tunable
coefficients. The linear function is $ax + b$, where $a$ and $b$ are tunable
coefficients.

\section{Results}
\label{section:results}

This section presents results using quantum circuit unoptimization for noise
scaling. One of the crucial components of zero-noise extrapolation is the
\emph{noise scale factor}. Here, we define the noise scale factor $\lambda$ as
the ratio between the total number of circuit operations before quantum circuit
unoptimization and after, see eq.~\eqref{equation:noise_scale_factor}. The total
number of circuit operations is the sum of the single and two-qubit gates, not
including qubit measurement or initialization operations. The intuition behind this
choice of scaling factor is that it follows the circuit complexity of unitary
circuit folding. However, this noise scale factor is only an
\emph{approximation}---the altered circuit structure can cause differences in
error propagation. 

\begin{figure*}[ht!]
    \centering
    \includegraphics[width=0.49\linewidth]{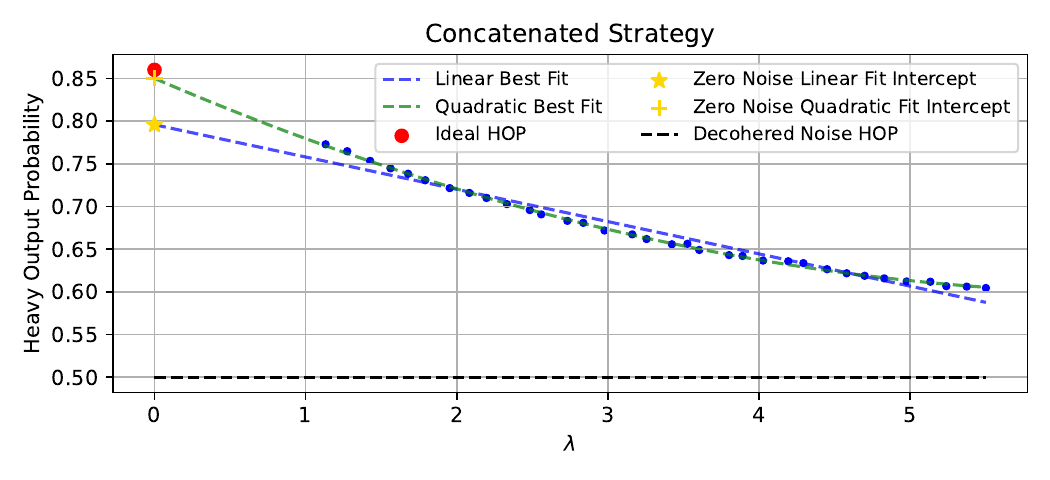}
    \includegraphics[width=0.49\linewidth]{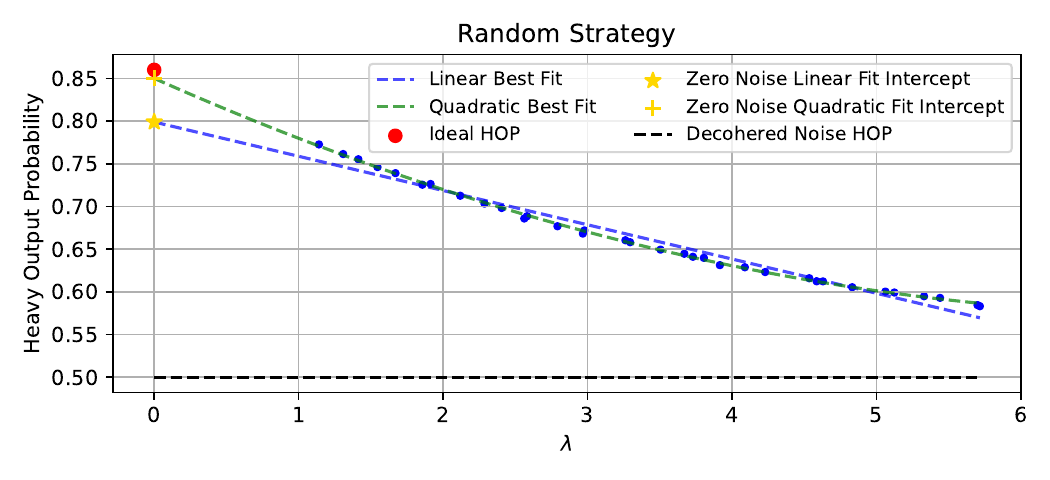}
    \caption{Zero-noise extrapolation using linear and quadratic regression for
    a Quantum Volume circuit with $10$ qubits using quantum circuit
    unoptimization gate selection method, random (right) and concatenated (left).
    There were $35$ circuit unoptimization iterations used to amplify the noise
    and each step's estimated observable is plotted as a blue point in each
    sub-figure. Each sub-plot shows the heavy output probability (y-axis) as a
    function of the noise scale factor $\lambda$. The red marker denotes the
    noiseless observable quantity, and the two yellow markers denote the quantum
    circuit unoptimization ZNE estimates of that observable. Note that each
    sub-plot used a single run of recursive quantum circuit unoptimization --
    meaning that the unoptimized circuit builds on itself as the number of steps
    increased, as opposed to independently choosing a new unoptimization path
    (with a new random seed) for each noise level.}
    \label{fig:ZNE_example_QV}
\end{figure*}

\begin{figure*}[ht!]
    \centering
    \includegraphics[width=0.49\linewidth]{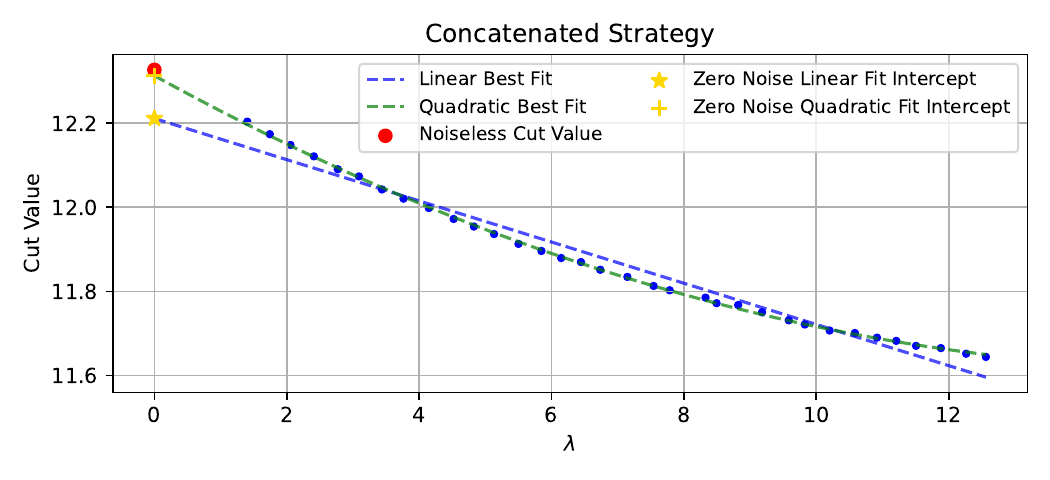}
    \includegraphics[width=0.49\linewidth]{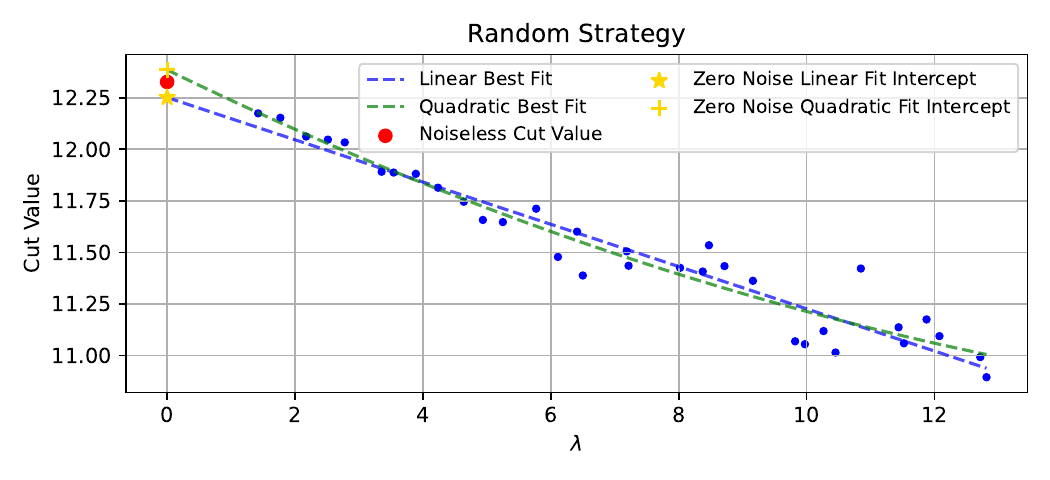}
    \caption{Zero-noise extrapolation using linear and quadratic regression for
    a fixed-angle $p = 2$ QAOA circuit applied to Max-Cut on a random 3-regular
    graph with $12$ qubits using quantum circuit unoptimization gate selection
    method random (right) and concatenated (left). $35$ quantum circuit
    unoptimization iterations were used to amplify the noise (the x-axis shows
    the computed $\lambda$ quantity for each one of these iterations), and each
    step's estimated observable is plotted in each sub-figure. Note that each
    sub-plot used a single run of recursive quantum circuit unoptimization --
    meaning that the unoptimized circuit builds on itself as the number of steps
    increased, as opposed to independently choosing a new unoptimization path
    (with a new random seed) for each noise level.}
    \label{fig:ZNE_example_QAOA}
\end{figure*}

The extrapolation functions we use in these simulations are only linear and
quadratic regression. This is because the observable scaling as a function of
$\lambda$ does not drop off according to an exponential function. However,
quantum-circuit unoptimization ZNE applied to other noise levels, quantum
circuits or observables could require other function fits, such as Richardson or
exponential extrapolation. 

Figure~\ref{fig:circuit_size_scaling} details gate count scaling and
correspondingly $\lambda$ scaling as a function of the number of quantum circuit
unoptimization iterations. This shows that quantum circuit unoptimization gives
a reasonably consistent, i.e., predictable, increase in circuit complexity.
However, the concatenated strategy gives a \emph{more} predictable scaling of
circuit gate operation count, whereas the random strategy gives more variable
gate operation counts.

Figure~\ref{fig:ZNE_example_QV} shows quantum circuit unoptimization ZNE applied
to a random quantum volume circuit. These results show that the noise can be
amplified predictably, and thus the noiseless signal can be recovered reasonably
well. 

Figure~\ref{fig:ZNE_example_QAOA} demonstrates using quantum circuit
unoptimization ZNE for a $p=2$ QAOA circuit applied to (unweighted) Max-Cut on a random
3-regular graph, where the observable is the graph cut value. These simulations used
fixed angles from ref.~\cite{wurtz2021fixed}. Notably, these results show a
clear difference between the two different two-qubit gate selection methods; the
random strategy, on average, results in more substantial noise amplification,
but is also more unpredictable, whereas the concatenated strategy results in a
more controlled and predictable noise amplification. The optimal Max-Cut value
for this specific problem instance is $16$ -- higher rounds of $p$ would need to
be used for QAOA to reach the optimal solution cut value.

\begin{figure*}[ht!]
    \centering
    \includegraphics[width=0.49\linewidth]{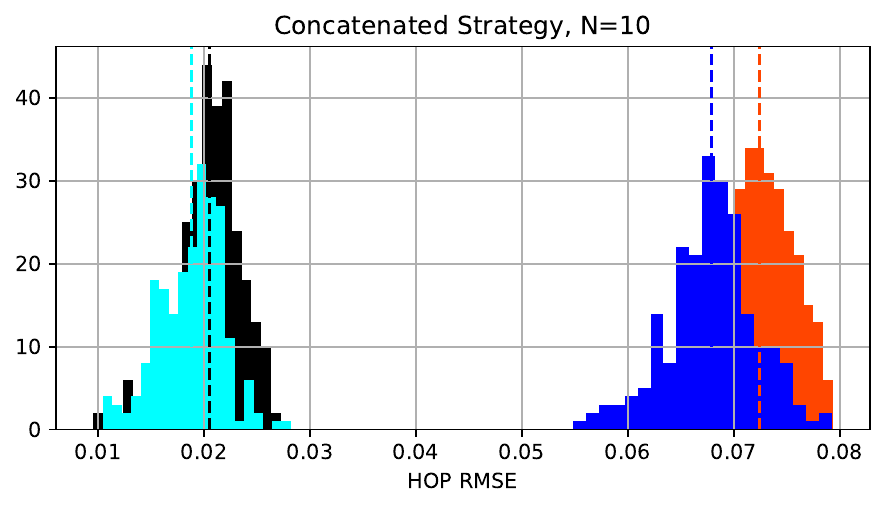}
    \includegraphics[width=0.49\linewidth]{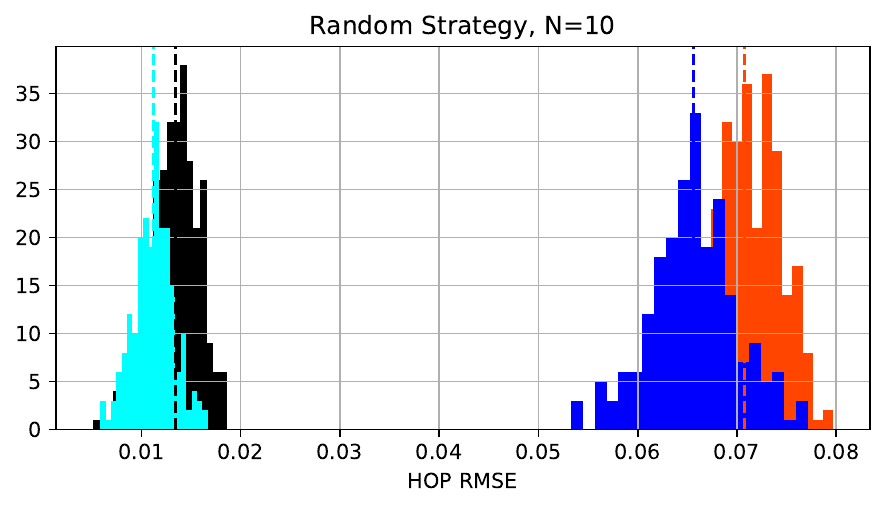}
    \includegraphics[width=0.58\linewidth]{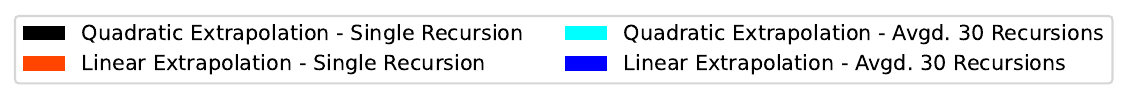}
    \caption{RMSE distributions between the ideal heavy output probability (HOP) and
    the ZNE extrapolated HOP using linear and quadratic extrapolation to
    mitigate the noise from a depolarizing error model when sampling a set of random $10$-qubit quantum volume circuits. Each distribution
    contains HOP results from a set of fixed $300$ randomly generated, and
    unique, quantum volume circuits. Each type of ZNE circuit unoptimization is
    applied to this set of random quantum volume circuits, and then, the two
    types of extrapolations are performed. Combined, these plots illustrate the
    differences in error rate between all combinations of the following 3
    circuit unoptimization ZNE parameters: linear vs. quadratic extrapolation,
    concatenated vs random strategy, and averaged over many circuit
    unoptimization passes vs. a single circuit unoptimization pass. The mean of
    each distribution is marked by the vertical dashed line corresponding to the
    color of that distribution. The overall lowest error rate set of parameters
    is quadratic extrapolation with the random two-qubit gate selection strategy
    and averaged over $30$ distinct recursions of circuit unoptimization (right
    plot, cyan distribution). Overall findings are as follows. Random
    strategy gives a lower error rate than concatenated strategy. Quadratic
    extrapolation performs better than linear extrapolation. Averaging the
    extrapolation over $30$ independent recursions of circuit unoptimization
    results in lower error rates than using only a single recursion run to
    extrapolate from.}
    \label{fig:HOP_RMSE_histogram}
\end{figure*}

Figure~\ref{fig:HOP_RMSE_histogram} reports distributions of heavy output
probability (HOP) root mean squared error (RMSE) from an ensemble of $300$, $10$ qubit, random quantum
volume circuits, when using quantum circuit unoptimization to perform ZNE to
mitigate the errors from a depolarizing error model. Lower RMSE corresponds to better approximation of the noiseless heavy output probability observable. Figure~\ref{fig:HOP_RMSE_histogram} additionally shows the
difference in error rate from using a single circuit unoptimization recursion
run compared to extrapolating based on many (in this case, $30$) independent
recursions. Here, independent recursion means explicitly that for each run we
have chosen a different random seed for the pseudorandom number generator in the
code to choose a slightly different path of two-qubit gates choices (only in the case of the random strategy), as well as
different random two-qubit unitaries $A$ (described in
Section~\ref{section:methods_quantum_circuit_unoptimization}). The result is
that performing noise averaging using the multiple circuit unoptimization
recursions results in lower error rates. This intuitively makes sense because
the circuit unoptimization procedure can cause many different circuit structures
with varying error propagation, and therefore averaging over many different
circuit structure cases will result in less biased observables. These results in
Figure~\ref{fig:HOP_RMSE_histogram} also show that the quadratic extrapolation
can approximate the correct heavy output probability observable better than
linear extrapolation. The concatenated gate selection strategy has a
higher error rate than the random gate selection strategy, which makes sense because the concatenated gate selection strategy does not use any random choice when choosing the pair of $2$-qubit gates, which means that the resulting observable extrapolation has less noise variance to average over.

\begin{figure*}[ht!]
    \centering
    \includegraphics[width=0.49\linewidth]{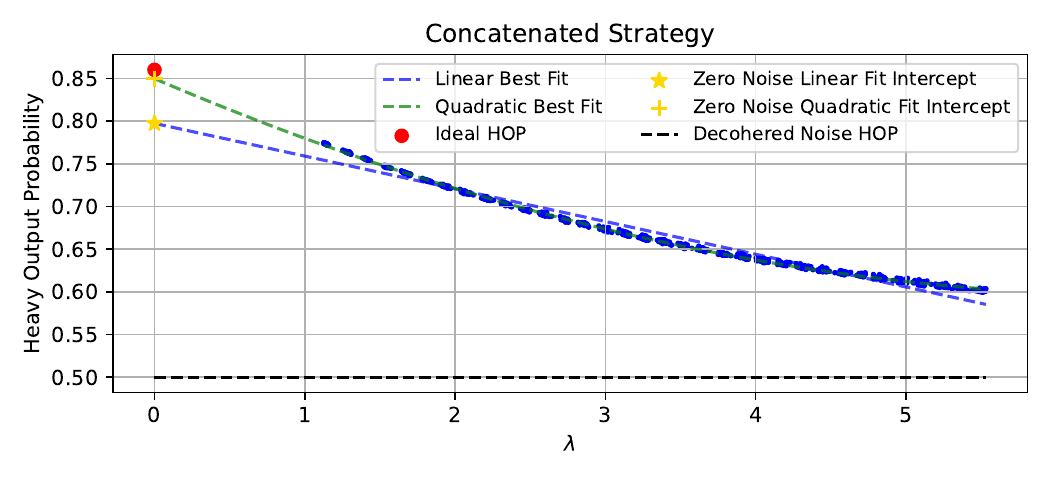}
    \includegraphics[width=0.49\linewidth]{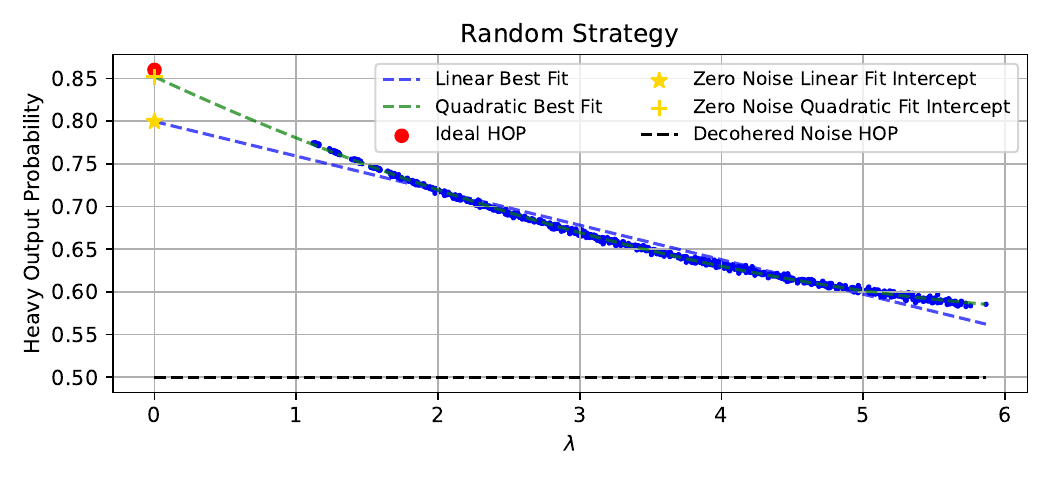}
    \includegraphics[width=0.49\linewidth]{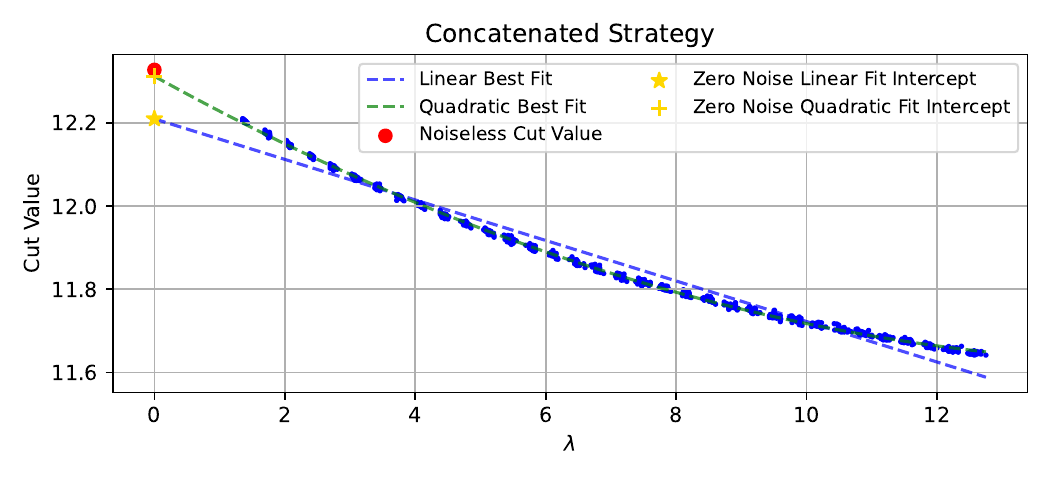}
    \includegraphics[width=0.49\linewidth]{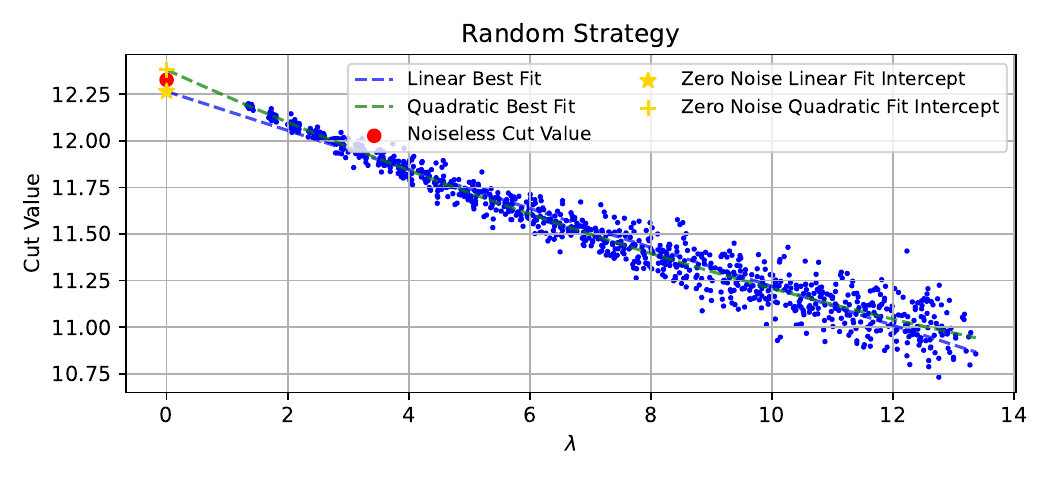}
    \caption{Averaged circuit variant zero-noise extrapolation numerical
    simulation tests. For each sub-plot, quantum circuit unoptimization is
    applied using a total of $30$ independent recursive sequences (each from $1$
    to $35$ iterations). This is unlike Figures~\ref{fig:ZNE_example_QV} and
    \ref{fig:ZNE_example_QAOA}, where only a single unoptimization sequence is
    used. The underlying quantum circuits (a fixed quantum volume circuit in the
    left column and a fixed Max-Cut QAOA circuit in the right-hand column), and
    observables, are the same as in Figures~\ref{fig:ZNE_example_QV} and
    \ref{fig:ZNE_example_QAOA}. Each separate quantum circuit unoptimization run
    can result in slightly different noise scale factors
    (eq.~\eqref{equation:noise_scale_factor}). The random selection strategy
    applied to the $p=2$ QAOA circuit has the highest observable variability.
    Still, in all four instances, the zero-noise extrapolation can recover a
    good signal from the noisy quantum circuit simulation. Notice that the variance seen in the individual data points for the concatenated strategy plots is purely due to the random two-qubit unitary selection during the insert stage of unoptimization. }
    \label{fig:ZNE_averaged}
\end{figure*}

Quantum circuit unoptimization with the random strategy allows some freedom in random selection of the
pair of two-qubit gates, one can generate many circuit variants (meaning they
implement equivalent unitaries but contain a distinct set of quantum instructions). Additionally, for both the concatenated and random strategy we can choose random two-qubit unitaries during the insert stage of unoptimization, which additionally results in many potential circuit variants. Figure~\ref{fig:ZNE_averaged} explores this by
repeating quantum circuit unoptimization on the same fixed test cases as
Figure~\ref{fig:ZNE_example_QV} and~\ref{fig:ZNE_example_QAOA}, but now running
$30$ complete recursions of the unoptimization (where each subsequent circuit is
constructed based on the gates created from the previous unoptimization run).
Each run used a new random seed so that the path of random gate selection and two-qubit unitary generation could
be different. We observe that some strategies and circuit types can yield highly
variable noise amplification, in this case, the random gate selection strategy
applied to the QAOA circuit instance, whereas the other three cases resulted in
more consistent noise amplification. Figures~\ref{fig:ZNE_averaged}
and~\ref{fig:HOP_RMSE_histogram} shows overall that the circuit averaging does
result in good observable recovery -- although notably the non-averaged cases
seen in Figure~\ref{fig:ZNE_example_QV} and \ref{fig:ZNE_example_QAOA} performed similarly. This shows that the average of extrapolated observables from many circuit variants can give robust
ZNE computation, but also that even single iterations of circuit unoptimization
still perform well. 

Combined, these results show that the concatenated strategy is better for
performing controlled (predictable) noise amplification using quantum circuit unoptimization,
but also results in higher error rates than the random gate selection option as
seen in Figure~\ref{fig:HOP_RMSE_histogram}, due to the lack of random $2$-qubit gate selection that can result in better noise averaging. Therefore, the best set of parameters that
result in the overall lowest error rate is to use the random gate selection
strategy, then average over many independent recursions of quantum circuit
unoptimization, and then to find a low-order polynomial extrapolation that gives the best curve fit to the observable that is being measured. 

\section{Discussion}
\label{section:discussion}

We explored using the quantum circuit unoptimization routines introduced
in~\cite{mori2024quantum} as a noise-scaling method for the zero-noise
extrapolation (ZNE) quantum error mitigation technique. To validate our
approach, we applied this noise-scaling method to random quantum volume and QAOA
circuits, using numerical simulations, to demonstrate its effectiveness.
These results show that quantum circuit unoptimization can provide controlled noise amplification while preserving the underlying logical unitary of the digital circuit.

An advantage of quantum circuit unoptimization compared to existing noise
amplification methods is that an exponentially growing number of distinct circuit variants can be generated for a particular
noise level. All other existing methods allow only one or a small number of
circuit variants to be generated for a given noise level. Generating many circuit variants facilitates better noise averaging, which could be specifically useful on
quantum computers with biased sources of error.

The primary limitation of quantum circuit unoptimization for noise amplification
in ZNE is the highly variable changes to circuit structure. These circuit
changes can result in very different error propagation compared to the
unmodified circuit. However, the previously mentioned property that \emph{many}
random (equivalent) circuit variants can be generated with quantum circuit unoptimization to mitigate this issue via noise
averaging across many instances with different circuit structures. Therefore, in
practice on noisy quantum computers, we recommend averaging over many random
circuits for each noise scale factor. A possible future subject of study is
developing other circuit unoptimization algorithms, in addition
to~\cite{mori2024quantum}, which create a quantum circuit structure that
minimizes highly biased error propagation (e.g., a noise-amplified circuit structure that approximately mimics the original circuit structure). 

It is worthwhile to consider the trade-offs of quantum circuit unoptimization as
a noise-scaling technique compared to other known techniques, such as unitary
folding. One of the benefits of quantum circuit unoptimization is that,
as more iterations of the unoptimization are performed, it becomes progressively
more difficult for the original circuit to be recovered by quantum computer backend compilers. This property is desirable because it ensures that
the noise amplification introduced by the unoptimization process remains intact
during execution, preventing compiler optimizations from inadvertently
simplifying the circuit. In contrast, unitary folded circuits can sometimes be
undone by the compiler, particularly when folding operations are explicitly
designed to be reversible and are thus more susceptible to simplification during
standard quantum circuit compiler optimization. By comparison, the structural complexity
introduced by unoptimization makes it more resistant to such optimizations. However, the original circuit structure is preserved in unitary folding, and the noise
amplification factor is more precisely determined, making it a more predictable
method in many scenarios. Note, however, that the circuit unoptimization routine could likely be identified and reversed by a compiler that was specifically designed to identify these circuit
structures -- it appears that current quantum circuit compilers are not able to
identify and reverse these structures~\cite{mori2024quantum}. Most current quantum computers 
provide options to disable backend compiler optimizations, mitigating the risk of
unwanted circuit simplifications. However, even when optimizations are turned
off, some low-level adjustments made by the compiler may still partially
reverse aspects of noise amplification methods used in ZNE, potentially leading to less
predictable noise amplification factors (for example, if the backend removes
identity gates that introduce errors due to idling time).

Using quantum circuit unoptimization means a slight change in how ZNE is
applied. Typically, when using ZNE, the user chooses a noise amplification
factor, and then the algorithm attempts to produce a circuit representation with
that level of noise. When using quantum circuit unoptimization,
instead, the noise scale factor (eq.~\eqref{equation:noise_scale_factor}) is
computed based on the circuits generated by the unoptimization, and that value
is then used for the extrapolation. Figure~\ref{fig:circuit_size_scaling} shows
that the circuit complexity scaling is predictable, which means that a noise
amplification factor in principle could be a priori requested. The closest integer 
required number of circuit unoptimization iterations could then be selected to
produce that noise amplification amount. However, this study used a more precise
measurement of the noise scale factor post-circuit unoptimization (including post Qiskit transpiler optimization). 

Another consideration is the relative complexity of implementation. Unitary
folding is conceptually simple and requires minimal additional overhead to
apply. By contrast, quantum circuit unoptimization involves multiple steps,
including gate insertion, swapping, decomposition, and synthesis, all of which
must be implemented carefully to ensure correctness. To partially offset this complexity, we provide a fully
self-contained implementation of quantum circuit unoptimization in Python~3,
compatible with IBM Qiskit circuits. This implementation can be adapted to other
quantum circuit libraries, ensuring accessibility for researchers and
practitioners interested in exploring unoptimization as a noise-scaling
technique.

Importantly, note that there is always a core assumption underpinning our method and
ZNE in general: the existence of a predictable relationship between the noise-scale factor and the measured expectation value, typically able to be modeled by a low-order polynomial.
Our results demonstrate that this
assumption holds in the numerical simulations we show, however it may break down for other problem classes, particularly near phase transitions or in algorithms with
unstable error propagation. It would be interesting for future work to
characterize classes of problems and noise models for which
quantum circuit unoptimization ZNE might break down.

Finally, it is essential to note that there is no theoretical
reason to expect quantum circuit unoptimization to outperform standard approaches for noise scaling such as unitary folding or pulse stretching. Each method has unique strengths and limitations, and the choice of technique depends on things such as the noise characteristics of the quantum hardware.

\section{Software}
\label{section:software}

The software implementing the quantum circuit unoptimization recipe, its
application as a noise-scaling technique for ZNE, as well as all data and plots
used in this work is available in a public GitHub repository
\begin{center}
    \scriptsize{\texttt{https://github.com/unitaryfund/circuit-unoptimization}}.
\end{center}

\section{Acknowledgments}
\label{section:acknowledgments}
The authors thank Yusei Mori for assisting in implementing the algorithm of
quantum circuit unoptimization and Andrea Mari, Farrokh Labib, Nate T. Stemen,
and Nathan Shammah for the helpful discussions on this approach. E.P. was
supported by the U.S. Department of Energy through the Los Alamos National
Laboratory the NNSA's Advanced Simulation and Computing Beyond Moore's Law
Program at Los Alamos National Laboratory. Los Alamos National Laboratory is
operated by Triad National Security, LLC, for the National Nuclear Security
Administration of U.S. Department of Energy (Contract No. 89233218CNA000001).
This work was partly supported by the U.S. Department of Energy, Office of
Science, Office of Advanced Scientific Computing Research, Accelerated Research
in Quantum Computing under Award Number DE-SC0020266 and Award Number
DE-SC0025336. This research used resources provided by the Darwin testbed at Los
Alamos National Laboratory (LANL), funded by the Computational Systems and
Software Environments subprogram of LANL's Advanced Simulation and Computing
program (NNSA/DOE). LANL report number LA-UR-25-21046.

\appendix
\section{Appendix}
\label{section:appendix}

In Figure~\ref{fig:unopt}, the circuit diagrams in the top panel may be
unfamiliar to readers who are not users of the Qiskit package. This section
provides an overview of the gates appearing in the diagrams and their
corresponding matrix representations. The standard set of single-qubit gates
used includes the Hadamard ($H$), Pauli-$X$, and Pauli-$Z$ gates, defined as
\begin{equation*}
    H = \frac{1}{\sqrt{2}}
    \begin{pmatrix}
        1 & 1 \\ 
        1 & -1
    \end{pmatrix}, \quad
    X = 
    \begin{pmatrix}
        0 & 1 \\ 
        1 & 0
    \end{pmatrix}, \quad
    Z = 
    \begin{pmatrix}
        1 & 0 \\ 
        0 & -1
    \end{pmatrix}.
\end{equation*}
The two-qubit gates used are the controlled-NOT ($CX$) and controlled-Z ($CZ$)
gates, defined as
\begin{equation*}
    CX = 
    \begin{pmatrix}
        1 & 0 & 0 & 0 \\ 
        0 & 1 & 0 & 0 \\ 
        0 & 0 & 0 & 1 \\ 
        0 & 0 & 1 & 0
    \end{pmatrix}, \quad
    CZ = 
    \begin{pmatrix}
        1 & 0 & 0 & 0 \\ 
        0 & 1 & 0 & 0 \\ 
        0 & 0 & 1 & 0 \\ 
        0 & 0 & 0 & -1
    \end{pmatrix}.
\end{equation*}
The $U$ and $U_3$ gates are equivalent in these circuit diagrams due to the
legacy naming convention from earlier versions of Qiskit, and represents the most
general single-qubit rotation. They are parameterized by three angles $\theta$,
$\phi$, and $\lambda$, and their matrix representation is given by
\begin{equation*}
    U(\theta, \phi, \lambda) = U_3(\theta, \phi, \lambda) = 
    \begin{pmatrix}
        \cos\frac{\theta}{2} & 
        -e^{i \lambda} \sin\frac{\theta}{2} \\
        e^{i \phi} \sin\frac{\theta}{2} & 
        e^{i(\phi + \lambda)} \cos\frac{\theta}{2}
    \end{pmatrix}.
\end{equation*}
The parameters $\theta$, $\phi$, and $\lambda$ determine rotations about the y-axis and z-axis of the Bloch sphere, making $U_3$ universal for single-qubit operations.

The $U_2$ gate is a special case of $U_3$ where the rotation angle is fixed at $\theta = \frac{\pi}{2}$. This simplifies the matrix to
\begin{equation*}
    U_2(\phi, \lambda) = U_3\left(\frac{\pi}{2}, \phi, \lambda\right) = 
    \frac{1}{\sqrt{2}}
    \begin{pmatrix}
        1 & -e^{i \lambda} \\
        e^{i \phi} & e^{i(\phi + \lambda)}
    \end{pmatrix}.
\end{equation*}
The $U_2$ gate can be considered as a simplified version of $U_3$ that performs
a Hadamard-like operation with additional phase shifts along the z-axis. Its
reduced parametrization makes it computationally efficient while retaining
sufficient flexibility for many quantum algorithms.

In addition to standard gates, some diagrams feature gates labeled `circuit-X'
(e.g., `circuit-826', `circuit-829'). These labels are automatically generated
by Qiskit when custom or composite gates are included in the circuit. A custom
gate represents a more complex operation that cannot be directly reduced to
standard gates like $U_3$ or $CX$. These custom gates typically correspond to
either composite sub-circuits representing a sequence of gates grouped and
treated as a single operation or as higher-dimensional unitary matrices applied
to multiple qubits.

\clearpage

\bibliographystyle{unsrtnat}
\bibliography{references}

\end{document}